\documentclass[aps,prd,superscriptaddress,nofootinbib,floatfix,preprint]{revtex4-1}

\usepackage[dvipdfmx]{graphicx}
\usepackage{amsmath}
\usepackage{bm}
\usepackage[small,bf]{caption}
\usepackage[subrefformat=parens]{subcaption}
\usepackage{comment}

\newcommand{\beq}{\begin{equation}}
\newcommand{\eeq}{\end{equation}}
\newcommand{\beqa}{\begin{eqnarray}}
\newcommand{\eeqa}{\end{eqnarray}}
\newcommand{\bpr}{\begin{problem}}
\newcommand{\epr}{\end{problem}}
\newcommand{\bcent}{\begin{center}}
\newcommand{\ecent}{\end{center}}
\newcommand{\bfig}{\begin{figure}}
\newcommand{\efig}{\end{figure}}
\newcommand{\bpc}{\begin{picture}}
\newcommand{\epc}{\end{picture}}
\newcommand{\barr}{\begin{array}}
\newcommand{\earr}{\end{array}}
\newcommand{\bitm}{\begin{itemize}}
\newcommand{\eitm}{\end{itemize}}
\newcommand{\bright}{\begin{flushright}}
\newcommand{\eright}{\end{flushright}}
\newcommand{\bminip}{\begin{minipage}}
\newcommand{\eminip}{\end{minipage}}
\newcommand{\btab}{\begin{tabular}}
\newcommand{\etab}{\end{tabular}}

\newcommand{\hiroshima}{Graduate School of Advanced Science and Engineering, Hiroshima University, Kagamiyama, Higashi-Hiroshima, Hiroshima 739-8526, Japan}

\newcommand{\icr}{Institute for Chemical Research, Kyoto University Uji, Kyoto 611-0011, Japan}
\newcommand{\tokai}{Research Institute of Science and Technology, Tokai University, 4-1-1 Kitakaname, Hiratsuka, Kanagawa 259-1292, Japan}

\newcommand{\kyoto}{Graduate School of Science, Kyoto University, Sakyouku, Kyoto 606-8502, Japan}
\newcommand{\ELINP}{Extreme Light Infrastructure-Nuclear Physics (ELI-NP)/Horia Hulubei National Institute for R\&D in Physics and Nuclear Engineering (IFIN-HH), 30 Reactorului St., P.O. Box MG-6, Bucharest-Magurele, Judetul Ilfov, RO-077125, Romania}
\newcommand{\NILPR}{National Institute for Laser, Plasma and Radiation Physics, 409 Atomistilor  PO Box MG-36, 077125, Magurele, Jud. Ilfov, Romania}

\newcommand{\om}{\omega}

\newcommand{\ve}{\varepsilon}

\newcommand{\Tsix}{\mathrm{T}^{6}}

\begin{document}
\title{Search for sub-eV axion-like particles 
in a stimulated resonant photon-photon collider with two laser beams
based on a novel method to discriminate pressure-independent components}

\author{Yuri Kirita}\affiliation{\hiroshima}
\author{Takumi Hasada}\affiliation{\hiroshima}
\author{Masaki Hashida}\affiliation{\icr}\affiliation{\tokai}
\author{Yusuke Hirahara}\affiliation{\hiroshima}
\author{Kensuke Homma\footnote{corresponding author}}\affiliation{\hiroshima}
\author{Shunsuke Inoue}\affiliation{\icr}\affiliation{\kyoto}
\author{Fumiya Ishibashi}\affiliation{\hiroshima}
\author{Yoshihide Nakamiya}\affiliation{\icr}\affiliation{\ELINP}
\author{Liviu Neagu}\affiliation{\ELINP}\affiliation{\NILPR}
\author{Akihide Nobuhiro}\affiliation{\hiroshima}
\author{Takaya Ozaki}\affiliation{\hiroshima}
\author{Madalin-Mihai Rosu}\affiliation{\ELINP}
\author{Shuji Sakabe}\affiliation{\icr}\affiliation{\kyoto}
\author{Ovidiu Tesileanu}\affiliation{\ELINP}
\collaboration{SAPPHIRES collaboration}

\date{\today}

\begin{abstract}
Sub-eV axion-like particles (ALPs) have been searched for by 
focusing two-color near-infrared pulse lasers into a vacuum along a common optical axis.
Within the focused quasi-parallel collision system created by combining a creation field
($2.5\,\mathrm{mJ}/47\,\mathrm{fs}$ Ti:Sapphire laser) and 
a background inducing field ($1.5\,\mathrm{mJ}/9\,\mathrm{ns}$ Nd:YAG laser),
the detection of signal photons via stimulated resonant photon-photon scattering by 
exchanging ALPs was attempted in a vacuum chamber.
The signal wavelength can be determined via energy-momentum conservation in the vacuum,
and it coincides with that determined from the atomic four-wave-mixing (aFWM) process.
In this work, the pulse energies were one order of magnitude higher than 
those in the previous search, allowing aFWM from optical elements to be observed 
as a pressure-independent background for the first time, 
in addition to the residual-gas-originating aFWM following a quadratic pressure dependence. 
In principle the four-wave-mixing process in vacuum via ALP exchanges (vFWM)
must also be pressure-independent, so the development of a new method for discriminating 
the optical-element aFWM is indispensable for increasing the pulse energies to 
the values needed for future upgraded searches. 
In this paper, we will present the established method for quantifying the yield from
the optical-element aFWM process based on the beam cross-section dependence. 
With the new method, the number of signal photons was found to be consistent with zero. 
We then successfully obtained a new exclusion region in the relation 
between ALP-photon coupling, $g/M$, and the ALP mass $m$,
reaching the most sensitive point $g/M = 1.14\times10^{-5}\,\mathrm{GeV^{-1}}$ 
at $m = 0.18\,\mathrm{eV}$.
\end{abstract}

\maketitle

\section{Introduction}
Axion and axion-like particles (ALPs) can be generic candidates for
the dark components of the Universe if their couplings to matter
are reasonably weak with the proper masses.
Axion is a pseudoscalar boson associated with the
breaking of the Peccei-Quinn symmetry 
which is introduced to solve the strong CP problem~\cite{PQ}, and could be cold dark matter.
Additionally, a {\it miracle} scenario~\cite{miracle} 
which explains both dark matter and inflation
predicts the existence of an ALP having a mass range that overlaps with the axion models.
Scalar fields, such as dilaton~\cite{dilaton,DEptp} and chameleon~\cite{chameleon,HuntingDE},
have been widely explored in attempts to explain the dark energy in the Universe.

In the low mass region, many experiments \cite{pvlas,alps,osqar,PTEP-EXP00} have been conducted focusing on ALP-photon coupling 
as a method of searching for ALPs.
In this paper, we focus on the coupling between sub-eV ALPs and two photons.
For scalar ($\phi$)- and pseudoscalar ($\sigma$)-type ALPs,
the following effective Lagrangians are assumed:
\beq
-L_{\phi} = gM^{-1}\frac{1}{4}F_{\mu\nu}F^{\mu\nu}\phi , \hspace{10pt} -L_{\sigma} = gM^{-1}\frac{1}{4}F_{\mu\nu}\tilde{F}^{\mu\nu}\sigma,
\label{eq_phisigma}
\eeq
where
$g$ is a dimensionless constant for a given energy scale $M$ at which
a relevant global symmetry is broken.
$F^{\mu\nu}=\partial^{\mu}A^{\nu}-\partial^{\nu}A^{\mu}$ is the electromagnetic
field strength tensor,
and its dual tensor is defined \footnote{We explicitly note that
the factor $1/2$ is newly multiplied to the definition used 
in the previous search~\cite{SAPPHIRES00}.  } 
as $\tilde{F}^{\mu\nu} \equiv \frac{1}{2}\ve^{\mu\nu\alpha\beta}F_{\alpha\beta}$ with
the Levi-Civita symbol $\ve^{ijkl}$.

\begin{figure}[!hbt]
\begin{tabular}{cc}
 \centering
  \includegraphics[keepaspectratio, scale=0.8]
  {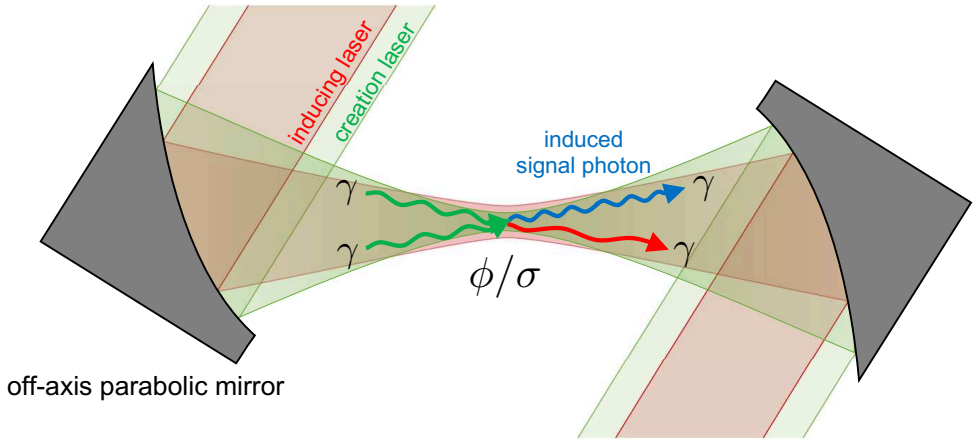}
\end{tabular}
\caption{Concept of a stimulated resonant photon-photon collision (vFWM process) via scalar ($\phi$) or 
pseudoscalar ($\sigma$) fields. The coaxially combined green and red lasers, which are, respectively, used 
to create ALPs and to simultaneously induce those decays, are focused with a parabolic mirror in a vacuum.
The wavy arrows represent two-body photon-photon scattering via ALP exchange in the laser fields.
The induced signal photons with wavelength determined by energy-momentum conservation (blue arrow) are 
produced primarily around the focal point.}
\label{fig:ccpt_SRPC}
\end{figure}

The SAPPHIRES collaboration searches for sub-eV ALPs 
via stimulated resonant photon-photon collisions (SRPC) by focusing two lasers in the manner illustrated in Fig.~\ref{fig:ccpt_SRPC}.
A collision between two photons with a shallow incident angle in a creation field (Ti:Sapphire laser, the green beam in Fig.~\ref{fig:ccpt_SRPC})
produces an ALP resonance state while a background field (Nd:YAG laser, the red beam in Fig.~\ref{fig:ccpt_SRPC}) simultaneously induces the ALP decay.
The wavelength of the signal photon (the wavy blue line) in the final state can differ from those 
of the lasers due to energy-momentum conservation.
In the terminology of laser physics,
this process can be interpreted as four-wave-mixing in vacuum (vFWM).

\begin{figure}[!hbt]
\begin{tabular}{cc}
 \centering
  \includegraphics[keepaspectratio, scale=0.8]
  {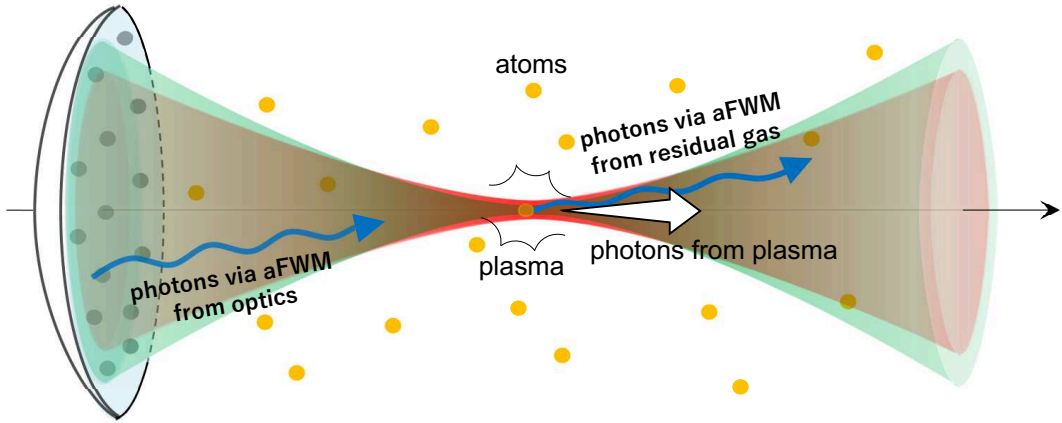}
\end{tabular}
\caption{Possible background sources originating from atoms and plasma 
in the quasi-parallel collision system. 
The atomic four-wave mixing (aFWM) photons can be produced from residual gas and optical elements, 
with the same wavelength as the signal photons. 
Plasma can be generated by focusing high-intensity laser fields at the focal point and ionizing the residual gas around it.}
\label{fig:Fig02}
\end{figure}

Figure \ref{fig:Fig02} illustrates possible background sources
with respect to vFWM. 
The dominant background process is four-wave mixing originating from 
atoms (aFWM). 
Both vFWM and aFWM are kinematically similar and indistinguishable
with the exception of polarization states of signal photons, which depend on the types of 
exchanged ALPs.
The dominant background source of aFWM is residual gas around 
the focal spot in the vacuum chamber that interacts with the high-intensity [$\mathrm{W/m^2}$]
laser fields.
However, a new background component originating from optical elements along the propagation path 
of the laser fields is expected to be significant for much higher intensities,
even when the beams have not been focused. 
In addition to aFWM, other background sources include the plasma generated 
by focusing lasers and the thermal noise in the photo-detection device. 

In the SAPPHIRES00 search~\cite{SAPPHIRES00}, 
we looked for ALPs with sub-mJ-level pulse lasers.
We provided a way to evaluate the background yields 
from residual-gas-originating aFWM, plasma, and other noise sources.
In this work, we used a $2.5\,\mathrm{mJ}/47\,
\mathrm{fs}$ Ti:Sapphire laser and 
a $1.5\,\mathrm{mJ}/9\,\mathrm{ns}$ Nd:YAG laser, 
whose energies are one order of magnitude higher than those in SAPPHIRES00.
Since the yield from optical-element aFWM was 
observed for the first time with mJ-level laser pulses, 
we have established a new method for discriminating between
the new background component and signals from vFWM. In the next section we summarize 
the principles behind the new method, which is based on the beam cross-section dependence.
We then present the new search results obtained using this method
and provide new exclusion regions for the relation between
ALP coupling with photons and ALP mass. Finally, we conclude with prospects for the future
work.

\section{Method for discriminating against optical-element atomic four-wave-mixing}
\label{sec_2}
To extract the yield from ALP exchanges, that is, vacuum-originating FWM 
$n_{vFWM}$, we first summarize the summation rules for the total number of signal-like FWM photons
$n_{FWM}$:
\beq\label{eq_nFWM}
n_{FWM} \equiv n_{vFWM} + n_{aFWM}
\eeq
with
\beq
n_{aFWM} \equiv n_{gas} + n_{opt},
\eeq
where $n_{gas}$ and $n_{opt}$ are the numbers of photons 
from residual-gas-originating aFWM and optical-element aFWM, respectively.
Another expression for $n_{FWM}$ based on the observed numbers 
of signal-like photons, denoted with capital characters as $N_{S,C,I,P}$, is given by
\beq\label{eq_nFWMobs}
n_{FWM} = N_{S} - (N_{C} - N_{P}) - (N_{I} - N_{P}) - N_{P}
\eeq
with
\beqa
N_{S} &=& n_{FWM} + n_{cPLS} + n_{iPLS} + n_{PED} \\
N_{C} &=& n_{cPLS} + n_{PED} \\
N_{I} &=& n_{iPLS} + n_{PED} \\
N_{P} &=& n_{PED},
\eeqa
where S, C, I, and P denote combinations of laser beams (see Fig.~\ref{fig:datataking_pattern} for details):
S for a signal-expected case with both creation and inducing beams, 
C for a case with only a creation laser, 
I for a case with only an inducing laser, and 
P for a case without beams, hence, a pedestal of the detection system.
$n_{cPLS}$ and $n_{iPLS}$ are the expected numbers of photons due to
plasma formation with only creation and inducing beams, respectively.

As had already been performed in past searches~\cite{PTEP2015,PTEP2020,SAPPHIRES00},
$n_{FWM}$ in Eq.(\ref{eq_nFWM}) can be determined from 
Eq.(\ref{eq_nFWMobs}) based on measurements in the different beam combinations above.
Past searches have also confirmed that
$n_{gas}$ follows a square dependence on the residual gas pressure. 
Thus, we first take a look at the pressure dependence of
$n_{FWM}$ by changing vacuum pressure values. 
If $n_{FWM}$ exhibits pressure independence,
then it implies that 
\beq\label{eq_const}
n_{FWM} = n_{vFWM} + n_{opt}
\eeq
in a low-pressure environment
because vFWM and optical-element aFWM are, in principle, pressure independent. 
Therefore, we need a method for evaluating $n_{vFWM}$ 
by discriminating against $n_{opt}$.

\begin{figure}[!hbt]
\begin{tabular}{cc}
 \centering
  \includegraphics[keepaspectratio, scale=0.7]
  {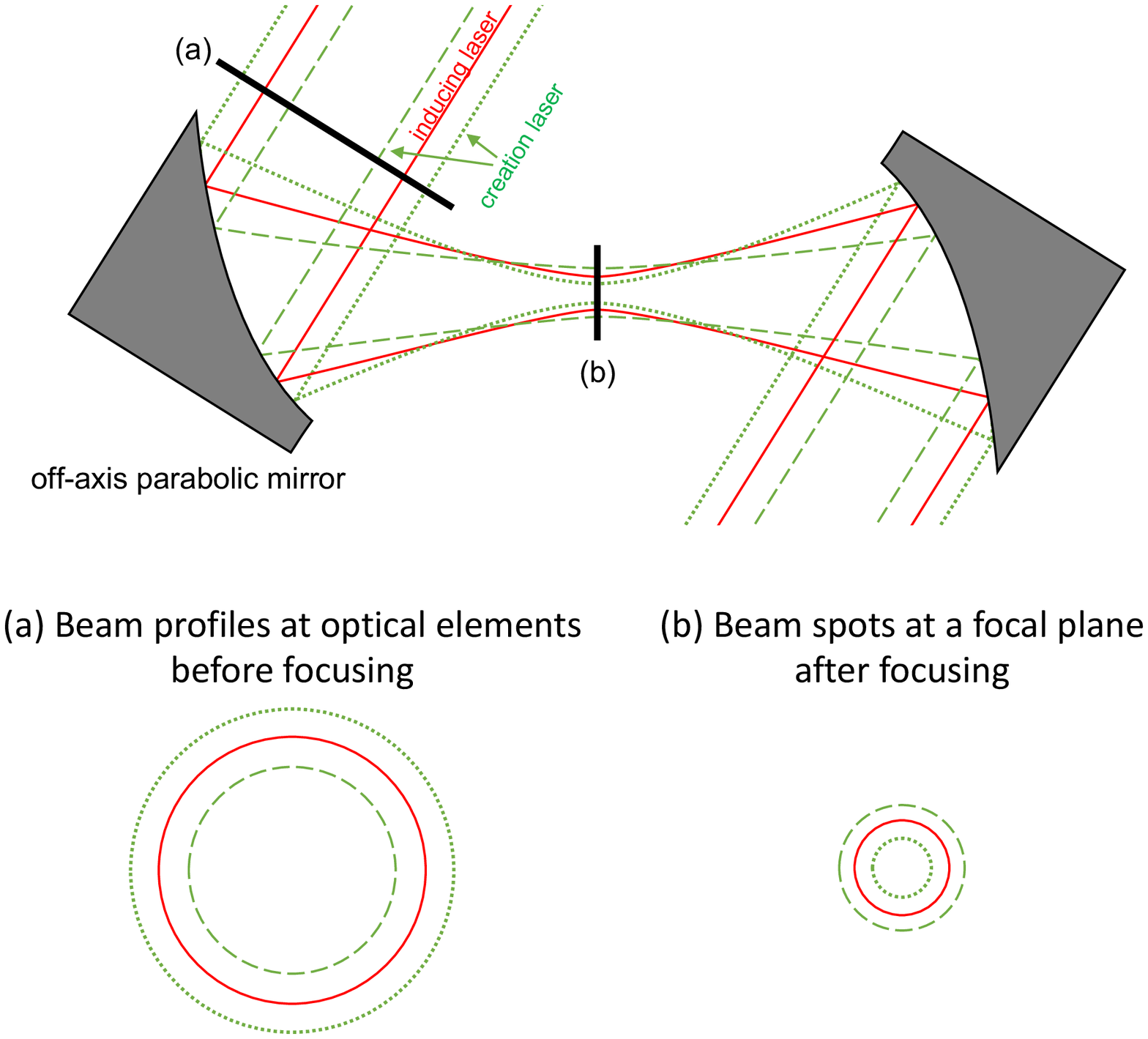}
\end{tabular}
\caption{Relations between beam cross sections for creation and inducing lasers: 
(a) at planes in optical elements before focusing and (b) at the focal plane after focusing.
(a) The creation and inducing laser cross sections are drawn with 
dotted and dashed green lines and a solid red line, respectively. 
The optical-element aFWM photons are produced at the overlapping areas between the cross sections of the creation and 
inducing beams.
(b) Behavior of the two beams at a focal plane after focusing is illustrated.
Since the beam waist is proportional to the inverse of the beam diameter at the focal point, a large (small) creation beam diameter
makes the beam waist small (large) and the beam intensity higher (lower).
Photons from vFWM and residual-gas-originating aFWM can be generated 
from the overlapping areas between the two beam cross sections,
while optical-element aFWM photons are never produced.
}
\label{fig:03}
\end{figure}

The principle for discrimination is illustrated in Fig.~\ref{fig:03}.
In any FWM process,
$n_{FWM}$ originates from the space-time overlapping region of the two laser pulses.
Figure \ref{fig:03}(a) shows cross sections of a creation laser (dotted and dashed lines)
with different beam diameters and an inducing laser (solid line) 
with a fixed diameter measured at one of the optical elements 
before focusing the two laser beams.
Figure \ref{fig:03}(b) shows the corresponding cross sections at a focal plane after focusing.
In Fig.~\ref{fig:03}(a), the peripheral ring beyond the inducing diameter 
(the area between the dotted and solid lines) will
never be able to produce aFWM because the inducing field does not exit there.
Therefore, when the creation beam diameter is beyond the range of 
the inducing beam diameter, as shown by the dotted line, optical-element aFWM yields
must be constant even if the creation beam diameter is increased.
On the other hand, in Fig.~\ref{fig:03}(b), since the beam waist $w_0$ 
in an ideal Gaussian beam at a focal plane is expressed as~\cite{Yariv}
\beq\label{eq_waist}
w_0 \equiv \frac{f\lambda}{\pi (d/2)}
\eeq
with laser photon wavelength $\lambda$, focal length $f$, and beam diameter $d$,
the dotted area is fully contained within the range of the inducing focal spot.
Increasing the creation beam diameter introduces a higher photon density at the
focal point, that is, a higher beam intensity in $\mathrm{W/m^2}$. Thus, both $n_{vFWM}$ and 
$n_{gas}$ are expected to be strongly dependent on the creation beam diameter.
Therefore, observing how $n_{FWM}$ increases 
when the creation beam diameter is increased, especially in the case where
the creation beam diameter is larger than that of the inducing beam,
enables discrimination against the optical-element aFWM.
If $n_{FWM}$ behaves in a pressure-independent fashion in 
a low-pressure environment, then we can test the dependence on 
beam cross sections to investigate whether the vFWM process is included or not 
on top of the constant behavior from optical-element aFWM. 
We must therefore know in advance how $n_{vFWM}$
behaves as a function of the beam cross sections.


\begin{figure}[!hbt]
 \begin{minipage}{0.48\hsize}
 \begin{center}
  \subcaption{$m = 0.1\,\mathrm{eV}$}\label{Fig04a}
  \includegraphics[keepaspectratio, width=80mm]
  {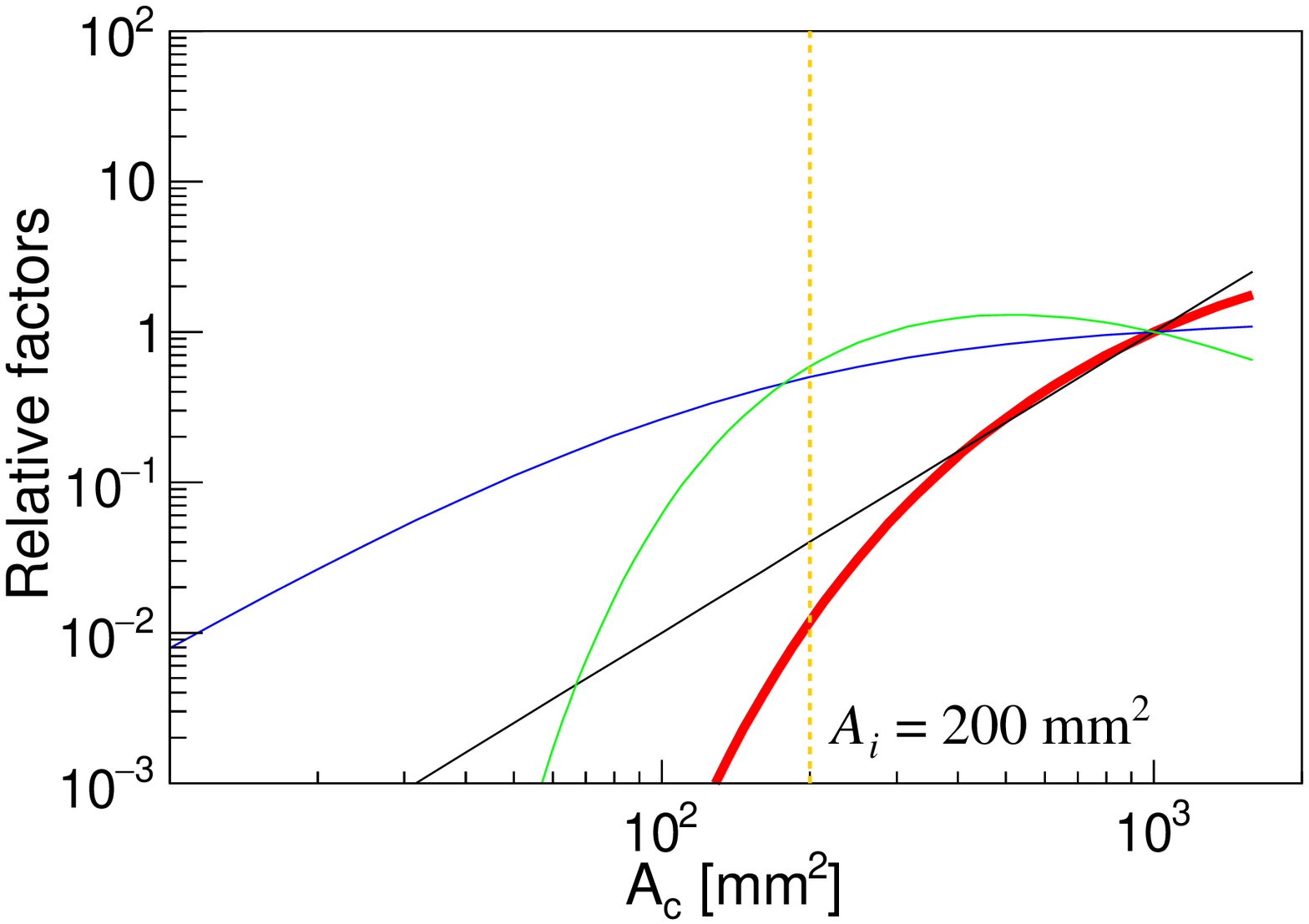}
 \end{center}
 \end{minipage}
 \begin{minipage}{0.48\hsize}
 \begin{center}
  \subcaption{$m = 0.01\,\mathrm{eV}$}\label{Fig04b}
  \includegraphics[keepaspectratio, width=80mm]
  {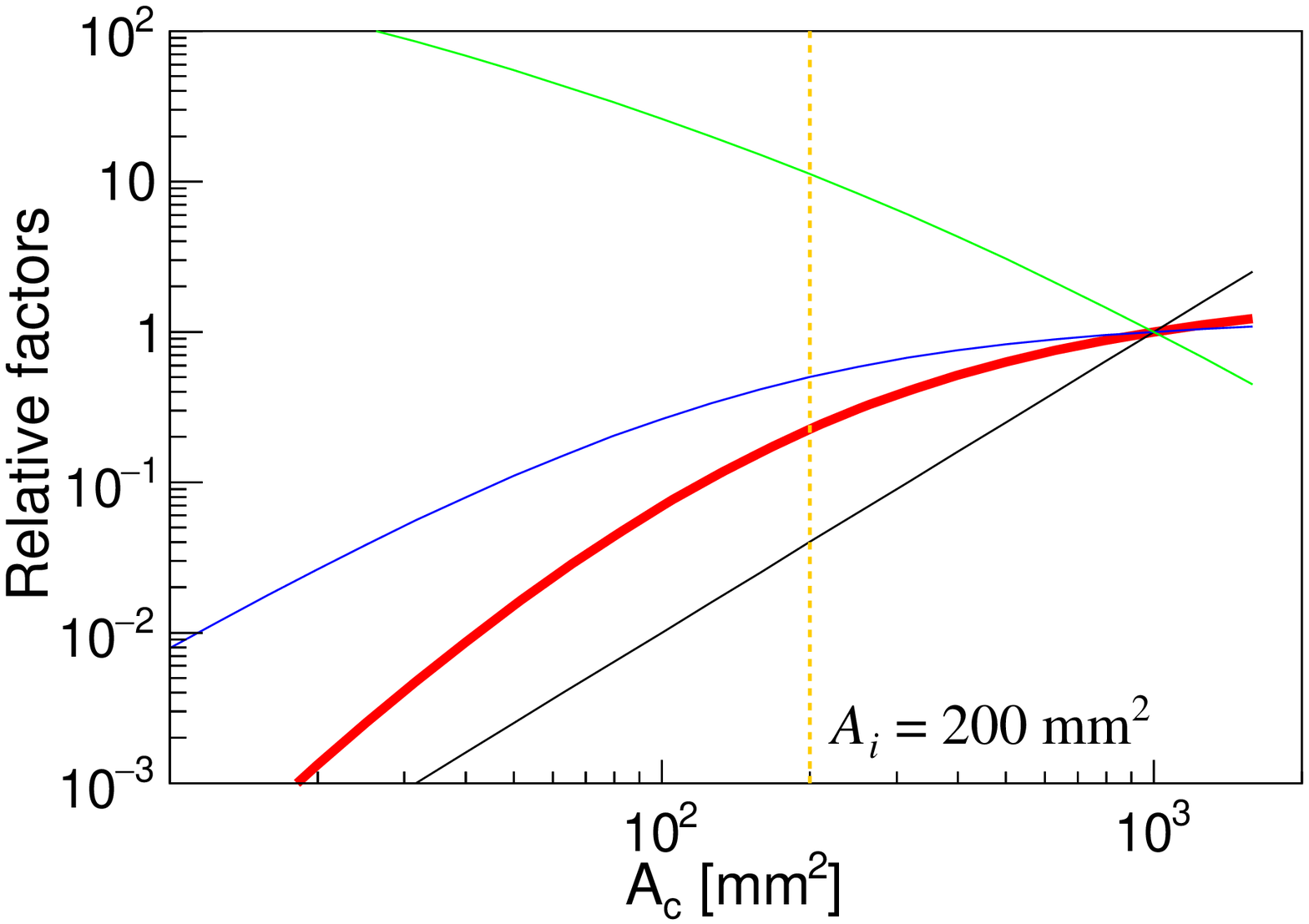}
 \end{center}
 \end{minipage}
 \begin{minipage}{0.98\hsize}
 \begin{center}
  \label{Fig04c}
  \includegraphics[keepaspectratio, width=160mm]
  {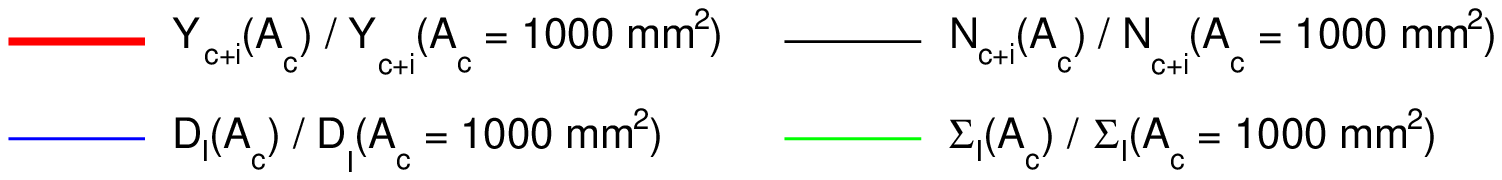}
 \end{center}
 \end{minipage}
 
 
\caption{Behavior of individual factors in Eq. (\ref{eq_Yci})
as a function of creation beam cross sections $A_c$
with respect to a fixed inducing beam cross section $A_i = 200\,\mathrm{mm}^2$ (yellow dashed line)
obtained by assuming an ALP mass of (a) $m=0.1\,\mathrm{eV}$ and (b) $m=0.01\,\mathrm{eV}$.
The red, black, blue, and green curves indicate $\mathcal{Y}_{c+i}$, 
$\mathcal{N}_{c+i} \equiv \frac{1}{4}N_c^2 N_i$, $\mathcal{D}_{I}$, and $\overline{\Sigma}_{I}$, 
normalized by their values at $A_c = 1000 \mathrm{mm}^2$, respectively.
Since the creation beam cross section is cut by an iris (A1 in Fig.~\ref{fig:Fig05_exp_setup}), the number of photons in a creation laser pulse $N_c$ is proportional to $A_c$.
$\overline{\Sigma}_{I}$, which describes ALP-photon interactions, has the characteristic beam 
cross-section dependence, because ALP resonance masses depend on the incident angles of 
two stochastically selected photons from among those in the focused creation beam.
}
\label{Fig04}
\end{figure}

The vFWM yield, which, more specifically from the particle physics point of view, 
is the signal yield per pulse collision from stimulated resonant photon-photon scattering, 
${\mathcal Y}_{c+i}$, is factorized as~\cite{JHEP2020,SAPPHIRES00}
\beqa\label{eq_Yci}
{\mathcal Y}_{c+i}[1] 
\equiv  \frac{1}{4} N^2_c N_i {\mathcal D}_I\left[s/L^3\right] \overline{\Sigma}_I\left[L^3/s\right].
\mbox{\hspace{2.3cm}}
\eeqa
$N_c$ and $N_i$ are the average number of photons in the two beams, 
with the subscripts $c$ and $i$ denoting the creation and inducing beams, 
respectively. ${\mathcal D}_I$ is a space-time overlapping factor between the incident beams,
and $\overline{\Sigma}_I$ is the {\it interaction volume rate}, not the
{\it interaction cross section}~\cite{JHEP2020,SAPPHIRES00}, specified in the units
within [\quad], which are expressed in terms of length $L$ and seconds $s$.

Figures \ref{Fig04} (a) and (b) demonstrate 
how individual factors in Eq. (\ref{eq_Yci}) 
behave as a function of the creation beam cross sections $A_c$
with respect to a fixed inducing beam cross section $A_i = 200\,\mathrm{mm}^2$, represented by the yellow dashed line.
The assumed ALP mass was $m=0.1$~eV and $m=0.01$~eV, respectively.
Individual factors are explained with different colors
inside the figures and have individual
normalizations of $A_c=1000$~mm${}^2$,
which is close to the actual value of the maximum creation beam cross sections used in this work.
What is important is the behavior of $\mathcal{Y}_{c+i}$ in the region $A_c > A_i$,
shown with the red curves.
Depending on the ALP mass,
$\overline{\Sigma}_I$ behaves differently, and the $\mathcal{Y}_{c+i}$ dependence
on $A_c$ obviously deviates from the constant behavior
seen when optical-element aFWM contributes as the sole background source.
Therefore, if $\mathcal{Y}_{c+i}$ exhibits a significant non-constant behavior for $A_c > A_i$ while in a low-pressure environment, then, in principle,
we can determine the mass scale of ALPs from 
the creation beam cross-section dependence~\cite{JHEP2020} 
in addition to testing the existence of ALPs.

\section{Experimental setup}
\label{sec_3}
\begin{figure}[b]
\begin{tabular}{cc}
 \centering
  \includegraphics[keepaspectratio, scale=0.8]
  {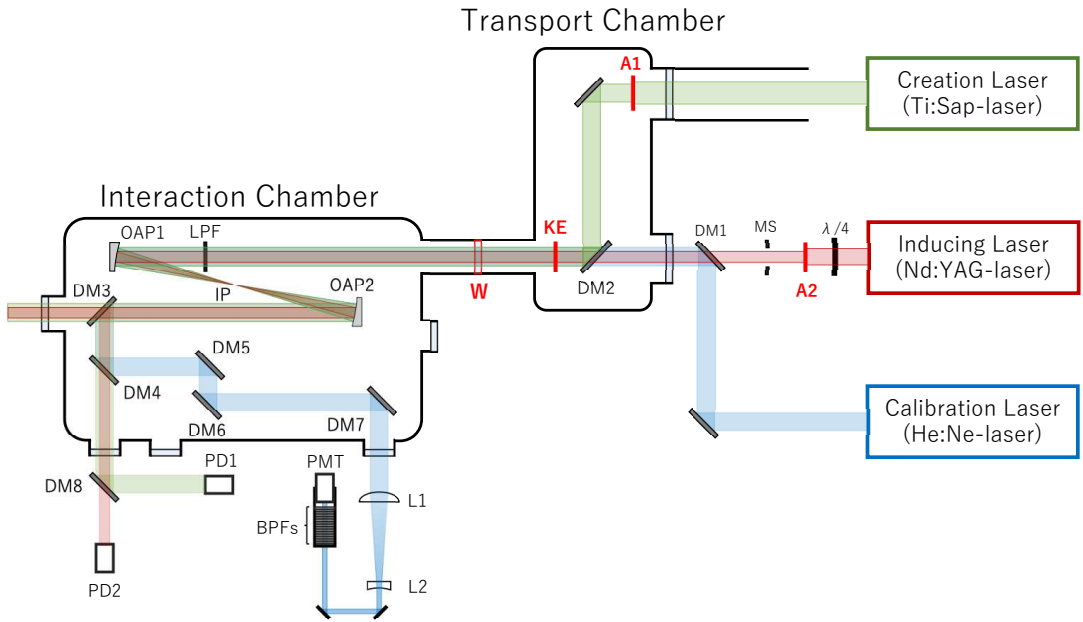}
\end{tabular}
\caption{Experimental setup for this work.
The basic design is the same as that in the previous search \cite{SAPPHIRES00}.
The propagations of individual lasers (creation laser, inducing laser, and calibration laser) are 
painted with green, red, and blue stripes, respectively.
Ti:Sapphire and Nd:YAG lasers were used as the creation and inducing lasers.
The calibration laser was a He:Ne laser, and it was used to obtain the acceptance 
of signal photons propagating from the Interaction point (IP) to the photomultiplier tube (PMT).
The two irises (A1 and A2) were introduced in order to change the creation beam diameter and to 
fix the inducing beam diameter for the measurement of the beam cross-section dependence.
Both beam diameters were determined using the knife-edge mechanism (KE).
We intentionally introduced the window (W)
not only to keep a low pressure value in the interaction chamber 
but also to enhance the background from optical-element aFWM, and 
to establish the method for discriminating against optical-element aFWM.
}
\label{fig:Fig05_exp_setup}
\end{figure}
Figure \ref{fig:Fig05_exp_setup} shows the experimental setup.
It is almost identical to the setup used in the previous search~\cite{SAPPHIRES00}.
For the creation and inducing fields we used a 47~fs-pulsed Ti:Sapphire laser, located 
at the Institute for Chemical Research in Kyoto University ($\Tsix$ system),
and a 9 ns-pulsed Nd:YAG laser.
The central wavelengths of these lasers were $816\,\mathrm{nm}$ and 
$1064\,\mathrm{nm}$, respectively.
The signal central wavelength was expected to be $662\,\mathrm{nm}$ via FWM:
$2\omega_c - \omega_i$ with creation photon energy $\omega_c$ and
inducing photon energy $\omega_i$.
A linearly polarized (P-pol) creation beam and a circularly polarized (left-handed)
inducing beam were combined at a beam combiner using a dichroic mirror (DM2)
in the transport chamber, which was maintained at $\sim 10^{-2}\,\mathrm{Pa}$.
The combined laser fields were focused coaxially into the interaction point (IP) 
in the interaction chamber, which was designed to reach $\sim 10^{-8}\,\mathrm{Pa}$
but was operated at $\sim 10^{-5}\,\mathrm{Pa}$ for this search.
Since the pressure values in the two chambers were so different, 
we introduced a window (W) between them. 
To reduce the background amount in the upstream so that the photo-detection device
could be operated without saturation, a set of long pass filters (LPF) for cutting 
the signal wavelength was inserted between the window and 
the focusing off-axis parabolic mirror (OAP1).
A second off-axis parabolic mirror (OAP2), identical to OAP1, was
placed at the symmetric position with respect to the IP to collect signal photons.
The signal photon selection was made through a set of five identical dichroic
mirrors (DM3 through DM7), and the collimated signal photons were
detected by a photomultiplier (PMT) through a set of tight band-pass filters (BPFs) that rejected
remnant beam photons. 
The voltage-time information was recorded with 
a waveform digitizer, and the peak finding algorithm was applied to 
obtain arrival times and the number of photons through amplitudes of waveforms.
The detailed specifications for these devices and for the peak analysis method can be
found in \cite{PTEP2020,SAPPHIRES00}.
To discuss the beam cross-section dependence of optical-element aFWM,
we introduced two irises (A1 and A2) with circular apertures along the individual beam propagations:
the c-Iris for changing the beam diameter of the creation laser 
and the i-Iris for fixing the beam diameter of the inducing laser.
To accurately determine the beam diameters before focusing, 
the knife-edge mechanism (KE) was installed after the DM2.

\section{Measurements}
\subsection{Arrival time distributions in four beam combinations}
\label{sec:sec4a_beamcombinations}
\begin{figure}[!t]
\begin{tabular}{cc}
 \centering
  \includegraphics[keepaspectratio, scale=0.5]
  {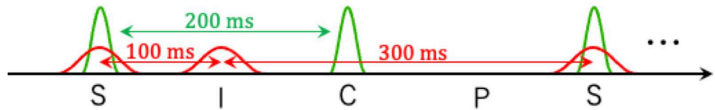}
\end{tabular}
\caption{Four beam combinations with the two laser pulses.
The green and red pulses are creation and inducing laser pulses, 
and the repetition rates are 5Hz at equal and irregular intervals, respectively.
We denote four pulse combinations:
S for both laser pulses,
C for only the creation laser pulses, 
I for only the inducing laser pulses, and
P for a pedestal measure without laser pulses.
}
\label{fig:datataking_pattern}
\end{figure}

To evaluate $n_{FWM}$ in Eq. (\ref{eq_nFWMobs}),
we configured four beam combinations with the two laser pulses.
The repetition rates of the creation and inducing laser pulses 
are 5Hz at equal and irregular intervals, respectively, 
as illustrated in Figure \ref{fig:datataking_pattern}.
\begin{table}[b]
\caption{Sources of photons included in individual beam combination patterns.
The characters $N_{S,C,I,P}$ denote the numbers of observed photons in the individual 
combination patterns expressed by the subscripts.
}
\begin{center}
\begin{tabular}{lcccc}  \\ \hline
Four beam combinations & $\ $S pattern$\ $ & $\ $C pattern$\ $ & $\ $I pattern$\ $ & $\ $P pattern$\ $ \\
Characters & $N_S$ & $N_C$ & $N_I$ & $N_P$ \\ \hline
Vacuum FWM ($n_{vFWM}$)& $\bigcirc$ & - & - & - \\
Atomic FWM ($n_{aFWM}$)& $\bigcirc$ & - & - & - \\ \hline
Plasma in the creation laser ($n_{cPLS}$)& $\bigcirc$ & $\bigcirc$ & - & - \\
Plasma in the inducing laser ($n_{iPLS}$)& $\bigcirc$ & - & $\bigcirc$ & - \\ \hline
Noise ($n_{PED}$)& $\bigcirc$ & $\bigcirc$ & $\bigcirc$ & $\bigcirc$ \\ \hline
\hline
\end{tabular}
\end{center}
\label{tab:KindsOfBG}
\end{table}
We then measured the arrival time distributions of photon-like signals
at the photomultiplier using peak finding on waveforms
for four pulse combinations: 
two laser pulses (Signal pattern, S), 
only the creation laser pulses (Creation pattern, C), 
only the inducing laser pulses (Inducing pattern, I), and 
without laser pulses (Pedestal pattern, P).
The expected sources of photons contained in individual patterns are 
summarized in table \ref{tab:KindsOfBG}.
The observed photons in the P pattern are pedestal components 
from all kinds of ambient noises, primarily from the thermal noise of the photo-detection device.
The observed photons in the C and I patterns contain 
plasma-originating photons from the individual lasers and pedestals.
The observed photons in the S pattern include signal photons and 
photons from all possible background sources.


\begin{figure}[!hbt] 
\begin{tabular}{cc}
 \centering
  \includegraphics[keepaspectratio, scale=0.7]
  {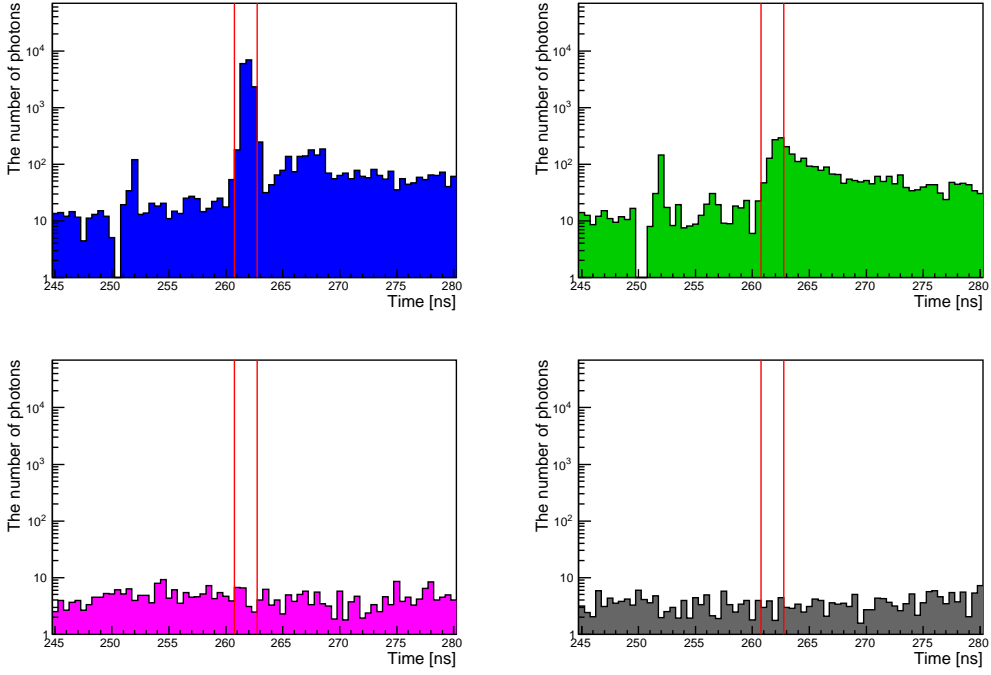}
\end{tabular}
\caption{Arrival-time distributions of observed photons in 
individual beam combinations at $4.32 \times 10^{-5}\,\mathrm{Pa}$. 
The red lines represent a time window within which 
signal photons are expected to arrive. 
The upper-left, upper-right, lower-left, and lower-right histograms correspond to the 
S, C, I, and P beam combinations, respectively.}
\label{fig:datataking_combination_graph}
\end{figure}

Figure \ref{fig:datataking_combination_graph} shows the arrival-time distributions 
of observed photons in individual combinations of laser beams.
The histograms for the S, C, I, and P patterns are shown in the upper left (Blue), 
upper right (Green), lower left (Purple) and lower right (Gray), respectively.
The area between the two red lines represents the expected arrival time window for signal photons.

\subsection{Pressure dependence of the number of FWM photons}

\begin{figure}[!hbt] 
\begin{tabular}{cc}
 \centering
  \includegraphics[keepaspectratio, scale=0.7]
  {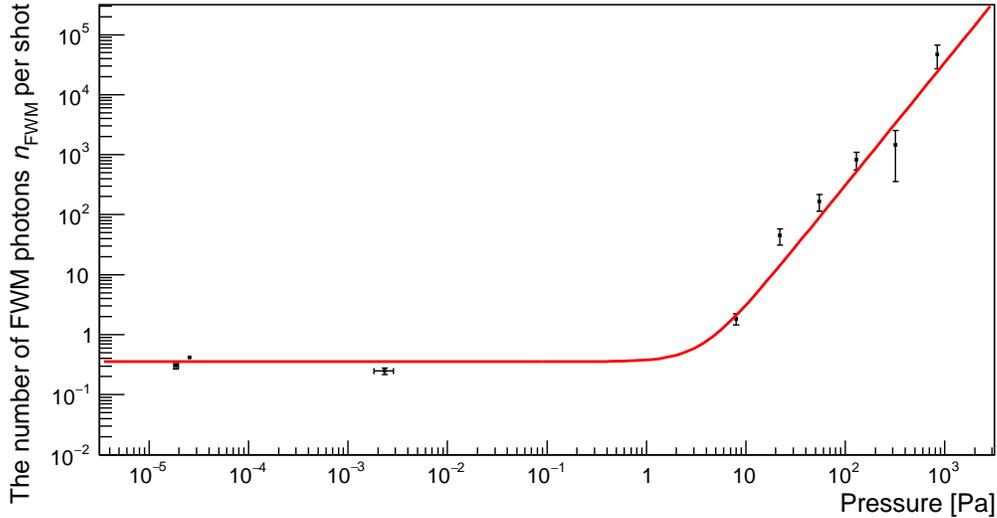}
\end{tabular}
\caption{Pressure dependence of the number of observed FWM photons $n_{FWM}$ per shot 
within the time window for expected signal arrival.
The number of observed FWM photons follows 
the square of the pressure when the chamber is above $10\,\mathrm{Pa}$.
This means that residual-gas-originating FWM backgrounds are dominant in high pressure environments.
Below $1\,\mathrm{Pa}$, a nonzero constant component was observed for the first time.
}
\label{fig:prsdep}
\end{figure}

Using subtractions between the four beam combinations based on Eq. (\ref{eq_nFWMobs}), 
we measured the number of FWM photons in the interaction chamber.
It is known that the amount of residual-gas-originating aFWM scales with the
square of the pressure values~\cite{PTEP2015,PTEP2020,SAPPHIRES00}.
Figure \ref{fig:prsdep} shows the pressure dependence of the number of 
FWM photons per shot $\mathcal{N}$, fit with the function
\begin{equation}
	\label{eq:pressure_dependence}
	\mathcal{N} = a+b\mathcal{P}^{c},
\end{equation}
where $\mathcal{P}$ is the pressure with fitting parameters $a$, $b$, and $c$.
The fit result, $c = 2.05 \pm 0.09$, is consistent with the square scaling.
From this result, we confirmed that the number of FWM photons via the residual gas
$n_{gas}$ was negligibly small at $10^{-5}\,\mathrm{Pa}$.
Additionally, the constant component was finally
observed as $a = 0.35 \pm 0.02$ for the first time.
\begin{figure}[!h] 
\begin{tabular}{cc}
 \centering
  \includegraphics[keepaspectratio, scale=0.7]
  {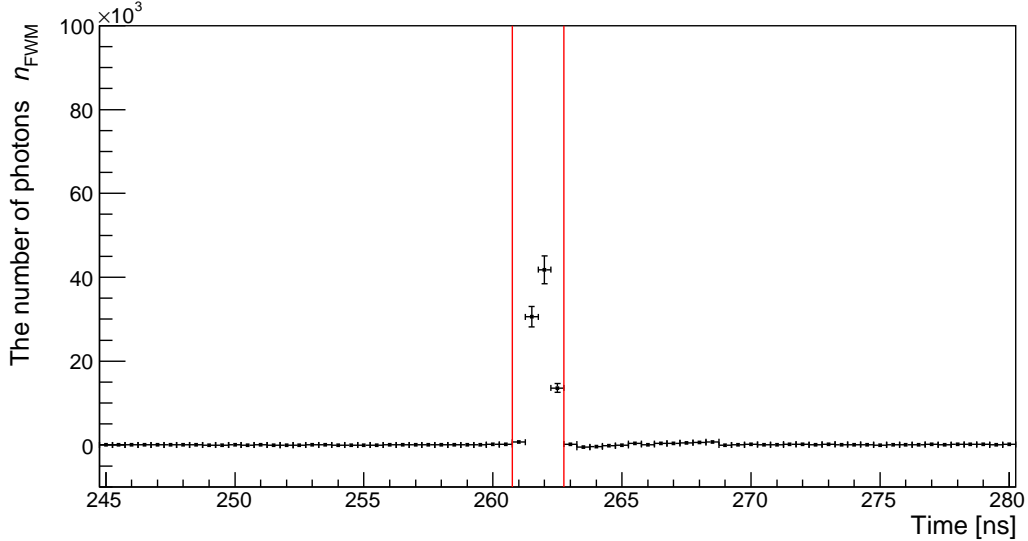}
\end{tabular}
\caption{Arrival-time distribution of the number of observed photons 
after subtracting background photons from plasma and thermal noise, 
based on Eq. (\ref{eq_nFWM}) with total S-pattern statistics of 15,000 shots at $2.87 \times 10^{-5}\,\mathrm{Pa}$.
The region between the two red lines is the time window within which signal photons are expected 
to arrive.
}
\label{fig:tansaku_dist}
\end{figure}
More explicitly, Fig.~\ref{fig:tansaku_dist} shows the arrival-time distribution
after the subtraction between the four beam combinations, based on Eq. (\ref{eq_nFWMobs})
with total S-pattern statistics of 15,000 shots at $2.87 \times 10^{-5}\,\mathrm{Pa}$.
This pressure-independent component is considered to be a possible mixture of
$n_{vFWM}$ and $n_{opt}$, as introduced in Eq. (\ref{eq_const}) of section \ref{sec_2}.

\subsection{Beam cross-section dependence of the number of observed photons}
\begin{figure}[!hbt] 
\begin{tabular}{cc}
 \centering
  \includegraphics[keepaspectratio, scale=0.7]
  {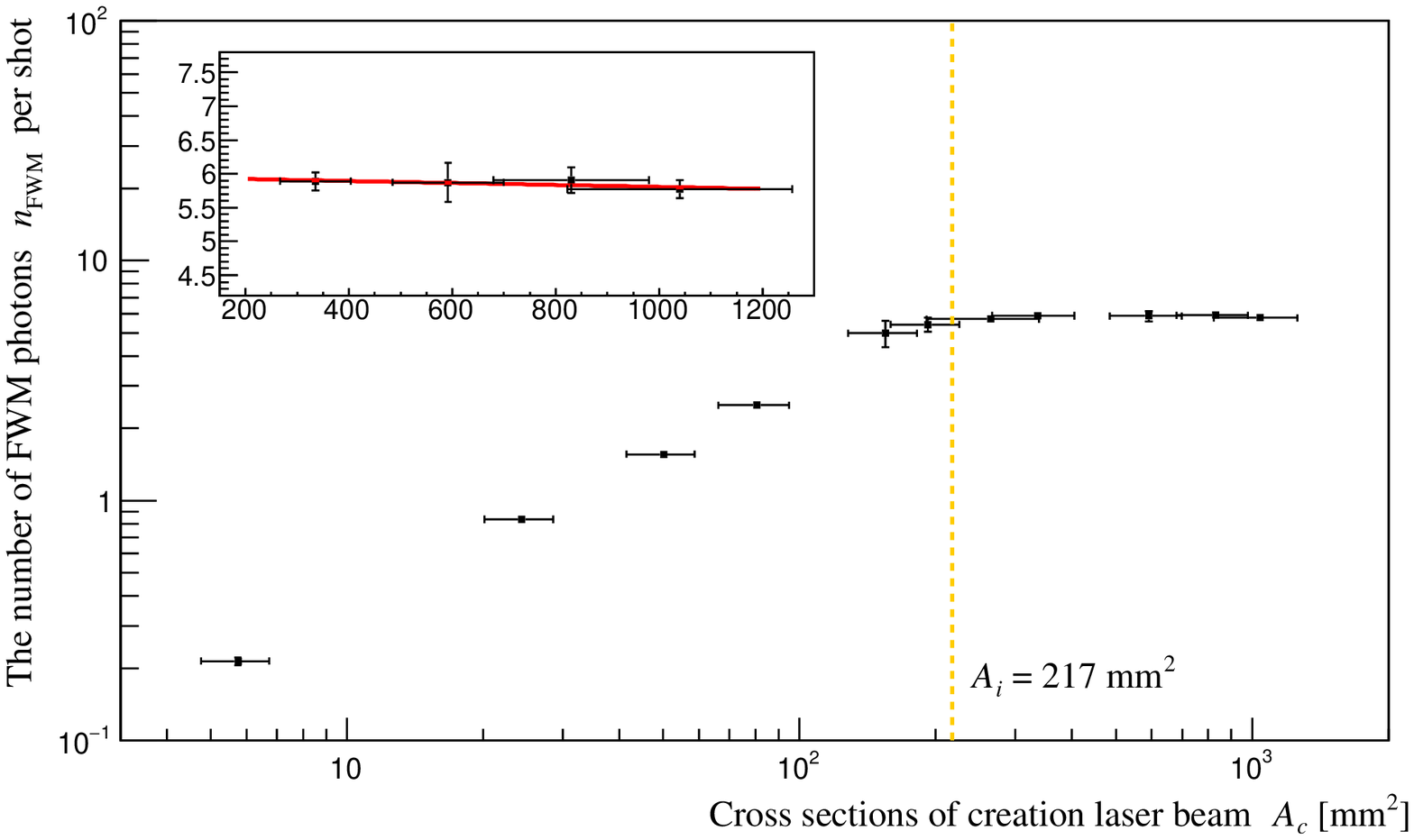}
\end{tabular}
\caption{Creation beam cross-section dependence of the number of observed FWM photons $n_{FWM}$ per shot, drawn on a double-logarithmic scale.
The yellow dashed line is a fixed inducing pulse cross section $A_{i} = 217 \,\mathrm{mm}^2$.
Four datapoints from the right-most point, where creation pulse diameters are sufficiently larger than the inducing pulse diameter with considering errors, are shown enlarged in the insert with a linear scale.
These were fitted with Eq. \eqref{eq_slope}, resulting in the slope $e = (-1.5 \pm 2.6) \times 10^{-4}$ represented by the red line.
Since the number of observed FWM photons are independent of the creation pulse cross section in the $A_c > A_i$ region, the optical-element aFWM accounts for the observed photons.
}
\label{fig:areascan}
\end{figure}
Figure \ref{fig:areascan} shows the beam cross-section dependence 
of the number of FWM photons at $10^{-5}$~Pa,
where the beam cross sections of the creation laser $A_c$ were changed with the variable size iris, c-Iris, 
while that of the inducing laser $A_i$ was fixed at $A_i = 217\,\mathrm{mm}^2$ (yellow dashed line in Fig.~\ref{fig:areascan}).
As discussed in section \ref{sec_2}, we only need to consider the dependence
on creation beam cross sections above the inducing beam cross section.
We thus focused on the last four cross section points,
and fitting was performed using the following linear function of the 
creation beam cross sections $A_c$ (red line in Fig.~\ref{fig:areascan}):
\begin{equation}\label{eq_slope}
	\mathcal{N} = d + e A_c.
\end{equation}
The slope $e$ was obtained as $e = (-1.5 \pm 2.6) \times 10^{-4}$.
Therefore, we conclude that the observed number of FWM photons is consistent
with being constant as a function of the creation beam cross sections, indicating that
the number of optical-element aFWM photons $n_{opt}$ occupies the $n_{FWM}$ in Eq. (\ref{eq_const}).

\section{Exclusion regions in ALP coupling-mass relations}
We obtained new exclusion regions in ALP coupling-mass relations
by following the same procedure as in the previous publication\cite{SAPPHIRES00}.
With a set of laser beam parameters $P$,
the number of ALP-originating signal photons, that is, vacuum-originating
FWM $n_{vFWM}$, as a function of mass $m$ and coupling $g/M$ is expressed as
\beq\label{Nobs}
n_{vFWM} = {\cal Y}_{c+i}(m, g/M ; P) t_{a} r \epsilon ,
\eeq
where $t_a$ is the data acquisition time, 
$r$ is the repetition rate of the pulsed beams, and
$\epsilon$ is the efficiency of detecting signal photons.
For a set of $m$ values and an $n_{vFWM}$,
a set of coupling $g/M$ is evaluated by solving this equation numerically.
The laser parameters are summarized in Table \ref{Tab2}.

\begin{table}[!ht]
\caption{Experimental parameters for numerical calculation of the exclusion limits in the $m - g/M$ parameter spaces.
}
\begin{center}
\begin{tabular}{lr}  \\ \hline
Parameter & Value \\ \hline
%
%
Creation pulse laser & \\
$\quad$ Centeral wavelength $\lambda_c$   & 816 nm\\
$\quad$ Relative linewidth, $\delta\omega_c/<\omega_c>$ &  $1.2\times 10^{-2}$\\
$\quad$ Duration time of pulse, $\tau_{c}$ & 40 fs \\
$\quad$ Measured pulse energy per $\tau_{c}$, $E_{c}$ & (2.53 $\pm$ 0.05) mJ \\
$\quad$ Pulse energy fraction within 3~$\sigma_{xy}$ focal spot, $f_c$ & 0.78\\
$\quad$ Effective pulse energy per $\tau_c$ within 3~$\sigma_{xy}$ focal spot & $E_{c} f_c = 1.97$~mJ\\
$\quad$ Effective number of creation photons, $N_c$ & $8.1 \times 10^{15}$ photons\\
$\quad$ Beam diameter of pulse, $d_{c}$ & (36.4 $\pm$ 2.2 $\pm$ 3.1)~mm\\
$\quad$ Polarization & linear (P-polarized state) \\ \hline
Inducing pulse laser & \\
$\quad$ Central wavelength, $\lambda_i$   & 1064~nm\\
$\quad$ Relative linewidth, $\delta\omega_{i}/<\omega_{i}>$ &  $1.0\times 10^{-4}$\\
$\quad$ Duration time of pulse, $\tau_{ibeam}$ & 9~ns\\
$\quad$ Measured pulse energy per $\tau_{ibeam}$, $E_{i}$ & $(1.51 \pm 0.02)$~mJ \\
$\quad$ Linewidth-based duration time of pulse, $\tau_i/2$ & $\hbar/(2\delta\om_{i})=2.8$~ps\\
$\quad$ Pulse energy fraction within 3~$\sigma_{xy}$ focal spot, $f_i$ & 0.80\\
$\quad$ Effective pulse energy per $\tau_i$ within 3~$\sigma_{xy}$ focal spot & $E_{i} (\tau_i/\tau_{ibeam}) f_i = 0.75$~$\mu$J\\
$\quad$ Effective number of inducing photons, $N_i$ & $4.0 \times 10^{12}$ photons\\
$\quad$ Beam diameter of pulse, $d_{i}$ & $(16.63 \pm 0.03 \pm 0.84)$~mm\\
$\quad$ Polarization & circular (left-handed state) \\ \hline
Focal length of off-axis parabolic mirror, $f$ & 279.1~mm\\
Single-photon detection efficiency, $\epsilon_{det}$ & 1.4\% \\
Efficiency of optical path from IP to PMT, $\epsilon_{opt}$ & 33\% \\ \hline
Total number of shots in S-pattern, $W_S$   & $15,000$ shots\\
$\delta{n}_{opt}$ & 427.7\\
\hline
\end{tabular}
\end{center}
\label{Tab2}
\end{table}

Since we conclude that the contribution from $n_{vFWM}$ is consistent with null 
from the creation beam cross-section dependence, 
we can evaluate new exclusion regions in the coupling-mass relation 
based on the fact that the background distribution
is only from the optical-element aFWM process. We do this by assuming the Gaussian
distribution on $n_{opt}$
because $n_{opt}$ is obtained from subtractions 
between different trigger patterns whose baseline fluctuations
are expected to individually follow Gaussian distributions.
We focus on the result from the highest ALP mass range, which corresponds to
$n_{opt}$ at the maximum creation beam cross section,
because a sensible mass $m$ is expressed as $m=2\omega_c \sin\vartheta$ 
with half the incident angles $\vartheta$ between two incident photons
varying in the range $0 < \vartheta < \sin^{-1}(f/(d/2))$.
A confidence level is thus defined with respect to the null hypothesis
by stating that the observed $n_{FWM}$ is all from the optical-element aFWM process.

A confidence level $1-\alpha$ for excluding the null hypothesis is expressed as
\beq
1-\alpha = \frac{1}{\sqrt{2\pi}\sigma}\int^{\mu+\delta}_{\mu-\delta}
e^{-(x-\mu)^2/(2\sigma^2)} dx = \mbox{erf}\left(\frac{\delta}{\sqrt 2 \sigma}\right),
\eeq
where $\mu$ is the expected value of an estimator $x$ corresponding to $n_{opt}$, and
$\sigma$ is one standard deviation $\delta n_{opt}$ on the measurement of $n_{opt}$.
We assign the acceptance-uncorrected uncertainty for $\delta n_{opt}$
from the quadratic sum of all error components in the following result
by multiplying the total statistics of 15,000 shots in the S pattern:
\beqa\label{eq_null}
n_{opt} = (8.6613 \pm 0.0427\mathrm{(stat.)} \pm 0.0023\mathrm{(syst.I)} 
           \pm 0.0009\mathrm{(syst.II)} ) \times 10^{4} \,\mathrm{photons}.
\eeqa
Systematic error I originates from baseline fluctuations
measured before the signal time window defined by the two red lines
in Fig.~\ref{fig:tansaku_dist}, reflecting the possibility of different pedestal distributions
depending on the different time domains.
Systematic error II comes from the threshold value used to define the single photon
waveform.
We note that the uncertainty 
of the focal spot overlap factor evaluated in \cite{SAPPHIRES00}
is omitted in this analysis because the optical-element aFWM process is 
insensitive to it, due to the the background only being produced when two parallel beams are propagating.

To obtain a confidence level of 95\%,
$2 \alpha = 0.05$ with $\delta = 2.24 \sigma$ was assigned, where
a one-sided upper limit by excluding above $x+\delta$~\cite{PDGstatistics} 
was applied.
For the evaluation of the upper limits on the coupling--mass relation,
we numerically solve
\beq
2.24 \delta n_{opt} = {\cal Y}_{c+i}(m, g/M ; P) t_{a} r \epsilon,
\eeq
where $t_a r = W_S = 15,000$, and
the overall efficiency $\epsilon \equiv \epsilon_{opt}\epsilon_{det} $,
with $\epsilon_{opt}$ being the optical path acceptance to the signal detector position
and $\epsilon_{det}$  the single photon detection efficiency.

\begin{figure}[!hbt]
	\begin{tabular}{cc}
		\centering
		\includegraphics[keepaspectratio, scale=0.7]{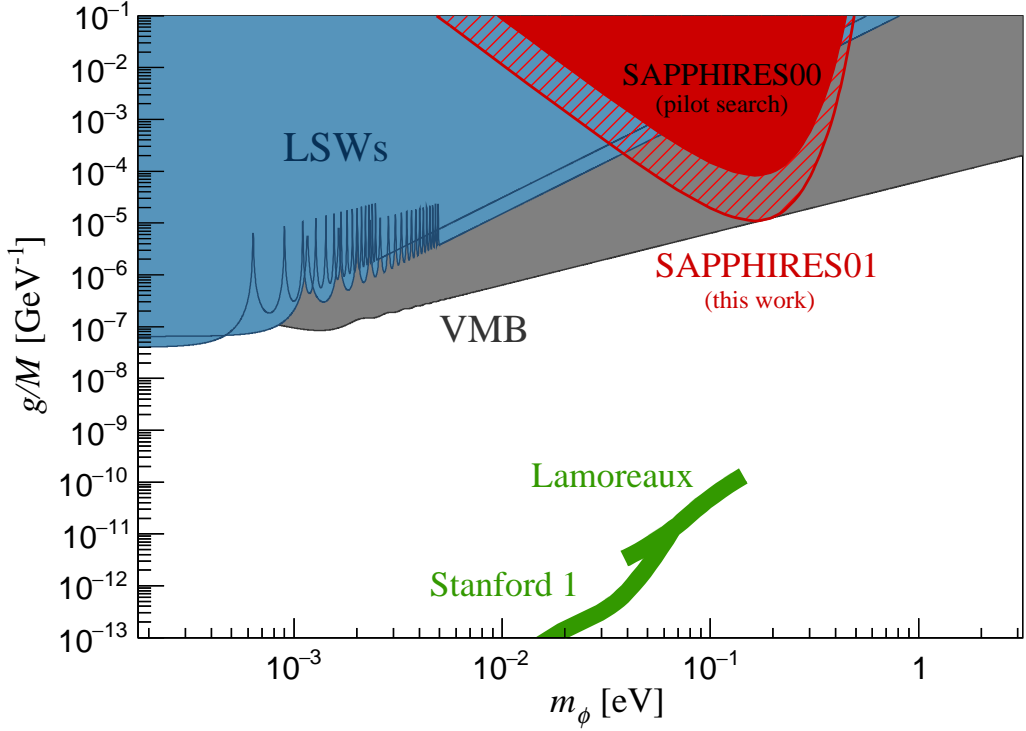}
	\end{tabular}
	\caption{
		Upper limit in the $m_{\phi}-g/M$ parameter space for scalar field exchanges achieved in 
this work, SAPPHIRES01, represented by the red shaded area. The red filled area is from the pilot search, SAPPHIRES00 \cite{SAPPHIRES00}. Limits from the LSW experiments (ALPS \cite{alps} and OSQAR \cite{osqar}) and the VMB experiment (PVLAS \cite{pvlas})  
are shown in the blue and gray areas, respectively. The green lines are exclusion limits from the non-Newtonian force 
searches (Stanford 1 \cite{Stanford1}) and the Casimir force measurement (Lamoreaux \cite{Stanford1}).
	}
	\label{fig:subeV_s}
\end{figure}

\begin{figure}[!hbt]
	\begin{tabular}{cc}
		\centering
		\includegraphics[keepaspectratio, scale=0.7]{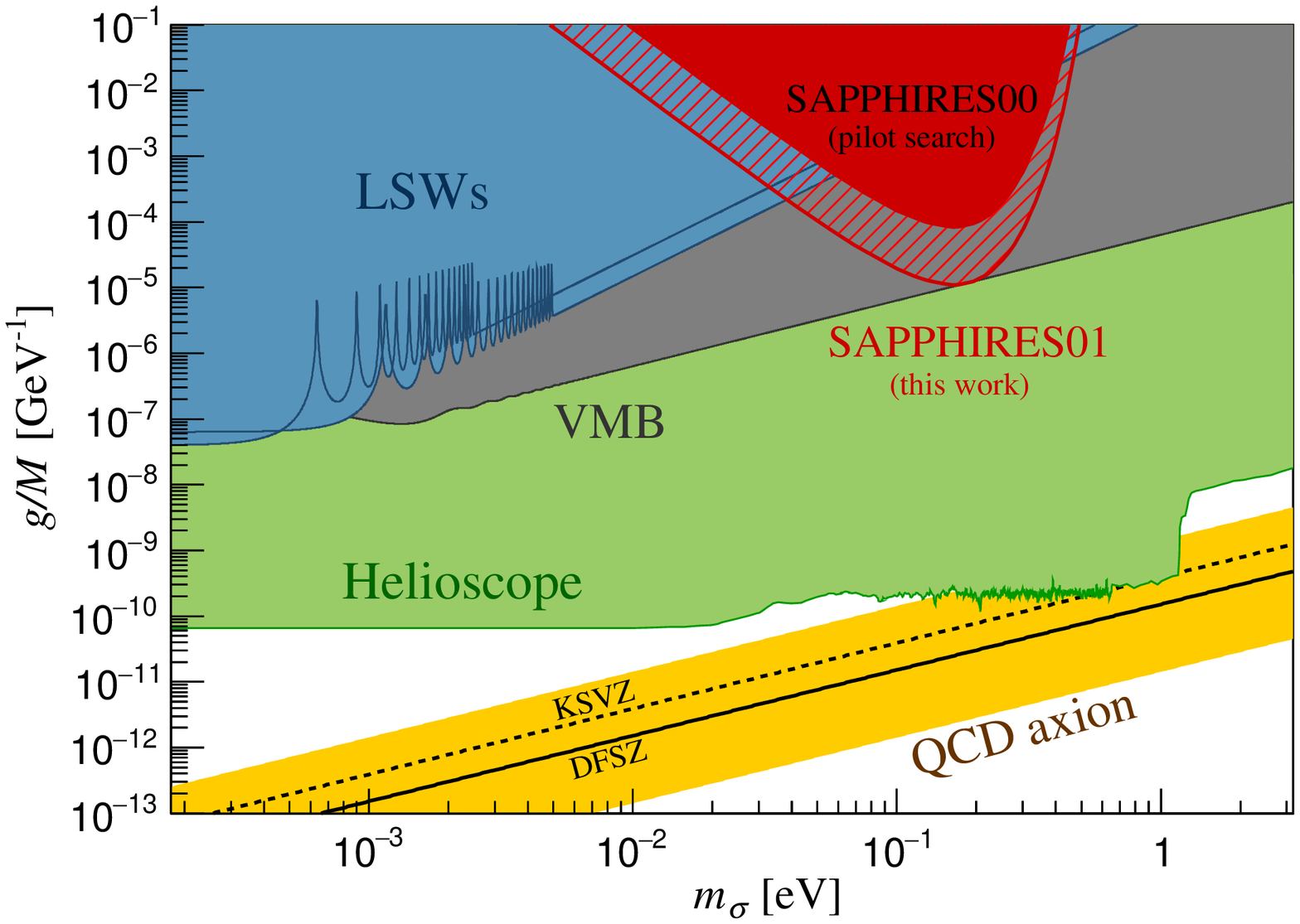}
	\end{tabular}
	\caption{
		Upper limit in the $m_{\sigma}-g/M$ parameter space for pseudoscalar field exchanges achieved 
in this work, SAPPHIRES01, represented by the red shaded area. 
The red filled area is from the pilot search, SAPPHIRES00 \cite{SAPPHIRES00}. 
Limits from the LSW experiments (ALPS\cite{alps} and OSQAR\cite{osqar}), 
the VMB experiment (PVLAS\cite{pvlas}), and a Helioscope experiment (CAST\cite{cast}) are shown 
with the blue, gray, and green filled areas, respectively. 
The yellow band and the black dashed lines represent QCD axion predictions 
from the KSVZ model \cite{KSVZ} with $0.07 < |E/N - 1.95| < 7$ and $E/N = 0$, respectively.
The solid line is the prediction from the DFSZ \cite{DFSZ} model with $E/N = 8/3$.
	}
	\label{fig:subeV_p}
\end{figure}

Figures~\ref{fig:subeV_s} and \ref{fig:subeV_p} provide the evaluated upper limits
of the coupling--mass relations
for scalar and pseudoscalar fields, respectively, at a 95\% confidence level.
The red shaded areas are from this work,
while the red filled areas are from previous searches \cite{SAPPHIRES00}~\footnote{
As for Fig.~\ref{fig:subeV_p}, we note that the limit from
the previous search \cite{SAPPHIRES00} is rescaled due to the factor 1/2 in the new definition
for the dual electromagnetic tensor in Eq. (\ref{eq_phisigma}).}.
The ALPS \cite{alps} and OSQAR \cite{osqar} experiments are ``Light Shining through Wall (LSW)'' experiments
and are filled with blue.
The individual sinusoidal terms in the sensitivities from ALPS and OSQAR are simplified to 1 above $5.0\,\mathrm{meV}$ and $2.5\,\mathrm{meV}$, respectively.
The gray area is the region excluded by the vacuum magnetic birefringence (VMB) 
experiment (PVLAS \cite{pvlas}). In Fig.~\ref{fig:subeV_s}, the lines with green are 
excluded by a non-Newtonian force search (Stanford 1 \cite{Stanford1}) and a Casimir force measurement (Lamoreaux \cite{Lamoreaux}).
In Fig.~\ref{fig:subeV_p}, the green area is the region excluded by a Helioscope 
experiment (CAST \cite{cast}). The yellow band is a region 
corresponding to the benchmark QCD axion model (KSVZ~\cite{KSVZ}) with $0.07 < |E/N - 1.95| < 7$.
The black dashed and solid lines are predicted
by the KSVZ model with $E/N = 0$ and the DFSZ \cite{DFSZ} model with $E/N = 8/3$, respectively.

\newpage
\section {Conclusions}
In this paper, we performed a stimulated resonant photon-photon collider experiment using two lasers with pulse energies around one order of magnitude higher than those 
in previous searches \cite{SAPPHIRES00}.
The generation of background photons via the atomic four-wave mixing (aFWM) process 
in optical elements was observed for the first time.
By measuring the creation beam cross section dependence of the number of 
FWM photons, we observed a constant behavior, indicating that
the observed photons originated only from
the optical-element aFWM background.
The expected value for the number of observed photons from resonant scattering 
via ALP exchanges was consistent with null. We then obtained 
the upper limit, which reached its maximum sensitivity in the coupling 
$g/M = 1.14\times10^{-5}\,\mathrm{GeV^{-1}}$ at $m = 0.18\,\mathrm{eV}$.

In this search the window material between the interaction chamber and 
the transport chamber was introduced to 
keep a high vacuum condition in the interaction chamber 
and to investigate how to handle optical-element aFWM background 
in real data analysis.
We then succeeded in establishing the method for discussing ALP exchanges under a nonzero pressure-independent background, based on the beam cross-section dependence
of the FWM yield.
In future searches a specialized intermediate chamber without the window material
will be inserted, which will reduce the optical-element background even if we increase the intensities of both beams. 
Building off of the established method,
we expect the upgrade to higher laser intensities to drastically improve the sensitivity 
to ALP exchanges.

\section*{Acknowledgments}
The $\Tsix$ system was financially supported by the MEXT Quantum Leap Flagship Program (JPMXS0118070187) and the program for advanced research equipment platforms (JPMXS0450300521).
Y. Kirita acknowledges support from the JST, the establishment of university fellowships for the creation of science technology innovation, Grant No. JPMJFS2129, and a Grant-in-Aid for JSPS fellows No. 22J13756 from the Ministry of Education, Culture, Sports, Science and Technology (MEXT) of Japan.
K. Homma acknowledges the support of the Collaborative Research
Program of the Institute for Chemical Research at Kyoto University 
(Grant Nos.\ 2018--83, 2019--72, 2020--85, 2021--88, and 2022--101)
and Grants-in-Aid for Scientific Research
Nos.\ 17H02897, 18H04354, 19K21880, and 21H04474 from the Ministry of Education, 
Culture, Sports, Science and Technology (MEXT) of Japan.
The authors in ELI-NP acknowledge the support from Extreme Light Infrastructure Nuclear Physics Phase II, 
a project co-financed by the Romanian Government and the European Union 
through the European Regional Development Fund and 
the Competitiveness Operational Programme (No. 1/07.07.2016, COP, ID 1334). 


\end{document}